\documentclass{article}

\usepackage[preprint]{neurips_2024}
\usepackage[utf8]{inputenc}
\usepackage[T1]{fontenc}
\usepackage{hyperref}
\usepackage{url}
\usepackage{booktabs}
\usepackage{amsfonts}
\usepackage{amsmath}
\usepackage{nicefrac}
\usepackage{microtype}
\usepackage{xcolor}
\usepackage{graphicx}
\usepackage{float}
\usepackage{array}
\usepackage{tabularx}
\usepackage{longtable}

\newcolumntype{L}[1]{>{\raggedright\arraybackslash}p{#1}}

\usepackage{colortbl}
\definecolor{effstrong}{HTML}{0F6E56}
\definecolor{effsolid}{HTML}{A85F0B}
\definecolor{effctx}{HTML}{6B6456}
\definecolor{phasebg}{HTML}{EEF4FB}
\definecolor{phasenote}{HTML}{5C574C}
\newcommand{\estrong}{\textcolor{effstrong}{\textbf{strong}}}
\newcommand{\estrongstar}{\textcolor{effstrong}{\textbf{strong*}}}
\newcommand{\esolid}{\textcolor{effsolid}{\textbf{solid}}}
\newcommand{\ectx}{\textcolor{effctx}{context}}
\newcommand{\ereq}{\textbf{required}}
\newcommand{\phaserow}[3]{\rowcolor{phasebg}\multicolumn{4}{l}{\textbf{#1.~#2}\quad\textcolor{phasenote}{\itshape #3}}\\ \midrule}

\arrayrulecolor{black!80}
\newcommand{\rowsep}{\arrayrulecolor{black!18}\midrule\arrayrulecolor{black!80}}

\title{The Effortless Trap: Productive Struggle, AI, and the Illusion of Learning}

\author{%
  Mario Brcic\textsuperscript{1,2} \qquad Stjepan Frljic\textsuperscript{1} \\
  \textsuperscript{1}Faculty of Electrical Engineering and Computing, University of Zagreb, Croatia \\
  \textsuperscript{2}It From Bit d.o.o., Zagreb, Croatia \\
  \texttt{Mario.Brcic@fer.hr}
}

\begin{document}
\maketitle

\begin{abstract}
\noindent
With AI advancing fast, educators face a dilemma: allow the tool or ban it. Conflicting evidence that it both helps and hurts learning only deepens the confusion. The allow-or-ban framing is a false dichotomy; the relevant design question is placement. Used well, AI can scale feedback, examples, practice, and individualized support. Used poorly, it replaces the cognitive work that learning requires and leaves an illusion of learning: a confident sense of mastery that collapses on the unaided task. The strongest causal evidence shows the outcome flips on design: an unguarded AI helper left high-school students about 17\% \emph{worse} on an unaided exam than peers with no tool at all, while the same model rebuilt to withhold answers erased the harm, and a well-engineered tutor roughly doubled learning. We give educators one graspable frame for placing the tool. A new idea is learned through six moves, in order: Prime, Probe, Point, Attach, Strengthen, and Test. Secure the first hard attempt and the final unaided check, scaffold with guarded AI in between, and one diagnostic carries the frame: if letting AI in makes the task feel effortless, it is in the wrong place. To make it usable, we map classical teaching moves and AI-supported interventions to each step. Together, the six-move model, the placement rule, and the intervention menu provide a practical foundation for lesson and course redesign in the age of AI.
\end{abstract}

\section{Introduction}

Every teacher now has to decide what to do about AI in their course, and most of the advice points the wrong way. The usual question is whether to allow it. That is the wrong question. The one that matters is \emph{where}, inside the act of learning, AI helps a student build a skill, and where it quietly does the work for them instead and harms learning.

It helps to start with how good teaching can get. For most of history the best education on earth was one devoted teacher beside one learner, and almost no one ever got it; it was reserved for princes and prodigies. Aristotle tutored Alexander of Macedon in person from about the age of thirteen; a decade later Alexander had conquered the known world. Helen Keller, cut off from language before she was two, reached a university degree after years of one-to-one teaching by Anne Sullivan. Closer to now, one small school in Budapest, the Fasori Lutheran Gymnasium, sent into the twentieth century John von Neumann and two future Nobel laureates, Eugene Wigner and John Harsanyi, its mathematics carried by masters like L\'aszl\'o R\'atz, who drilled von Neumann and Wigner before he named a method. Closer still, one informatics teacher in Gdynia, Ryszard Szubartowski, coached his students to sixty international-olympiad medals, several of whom went on to build OpenAI \citep{mistrz2025}. These examples are not the evidence, but they set the bar: they show what unusually patient, individualized, demanding teaching can do when it reaches the right learner at the right time. That is why AI matters for education: it may make such conditions less rare.

The historical examples above are exceptional, but the mechanism behind them is not mysterious. The power of one-to-one teaching is not only inspirational; it is empirically known to work. A good human tutor lifts an ordinary student well beyond the average classroom; the honest size of that effect is about $d \approx 0.79$, large but short of the celebrated two sigma \citep{vanlehn2011,bloom1984}. The trouble has always been supply. Not every student can have a R\'atz beside them every day; more importantly, not every teacher has had the time, tools, and support to offer that kind of attention at scale. This is where AI changes the arithmetic. The patient one-to-one attention that was scarce can, in principle, be made available more cheaply, for more students, more often. Early work points that way: a human-AI system raised the quality of real tutors at about \$20 per tutor per year \citep{wangcopilot2024}. We state this as promise, not delivered fact. The AI-specific evidence is young, and the same design space contains both the tutoring upside and the failure modes.

But AI is not a magic solution. In one careful randomized trial of nearly a thousand high-school mathematics students, the design of the tool, not the presence of the tool, decided the outcome \citep{bastani2025}. Students in the unguarded-AI group, where the tool behaved like an ordinary chatbot, performed better while they had it during practice but then scored about 17\% \emph{worse} than the no-access control group on the exam they sat alone. Students in the guarded-AI group used the same underlying model, rebuilt to \emph{withhold answers} and offer teacher-designed hints; that guardrail essentially erased the harm. The same underlying model produced opposite learning outcomes because its role was redesigned. This distinction matters: the practice-time gains and the unaided-exam loss came from different versions of the tool, not from one uniform intervention. The lesson is narrow and precise. Answer-giving where the student should struggle is what harmed learning, while guarded design in the same place did not. The upside is just as design-dependent. In a separate randomized study of undergraduate physics, a carefully engineered AI tutor produced more than twice the learning of a strong active-learning class \citep{freeman2014,crouchmazur2001}, in less time \citep{kestin2025}. So the practical choice is not allow-or-ban. It is where the tool sits in the learning, and what job it is asked to do there. The rest of this paper is about telling those cases apart.

Why should placement decide the outcome at all? Because learning happens only when the student does the effortful part, the wrestling that cognitive science calls \emph{productive struggle} \citep{kapur2008,sinhakapur2021}. An AI that does that part for the student feels like help while removing the very work that builds the skill \citep{risko2016}, and it does something more insidious: it feels like learning while it happens. The felt sense of learning is a poor gauge of the fact of it. Students in effortful active classes judged that they were learning less even as they objectively learned more \citep{deslauriers2019}, so fluent, AI-smoothed practice can feel most productive exactly when it teaches least. This is the \emph{effortless trap}, and what it leaves behind is an \emph{illusion of learning}: a confident sense of mastery that collapses on the unaided task. A controlled study makes the disconnect concrete: software engineers given an AI assistant on a coding task performed it well but learned markedly less, scoring far lower afterward on the concepts behind the code they had just produced \citep{shen2026}. The diagnostic that follows is blunt, and we return to it throughout: if letting AI in makes the task feel effortless, it is in the wrong place.

The contribution is a placement frame, not a curriculum. We describe how a student learns one idea as six moves (Section~\ref{sec:model}), state one rule for where AI belongs (Section~\ref{sec:rule}), and lay out a set of options, classical and AI-supported, mapped to those moves (Section~\ref{sec:ideas}). The aim is to give educators a practical way to decide when AI should support learning, when it should be constrained, and when it should stay out. We place the frame among related work in Section~\ref{sec:related} and state its limits plainly in Section~\ref{sec:limits}. Full course redesign requires further machinery; this paper supplies the placement decision that must come first.

\section{Related work}
\label{sec:related}

Recent work is converging in the same direction: AI use has to be designed around cognitive role, timing, and constraints. Assessment-oriented frameworks make this shift at the level of assessment design. The AI Assessment Scale, in its original and revised forms, gives a tiered vocabulary from no-AI to full-AI use, grounded in assessment redesign \citep{furze2024aias,perkins2025aias}. The University of Sydney two-lane model separates secured assessment of learning from open assessment for learning \citep{sydney2024,dawson2021,bearman2024}. Developmental and instructional frameworks make a parallel move from the learner and task side: a five-tier framework calibrates AI exposure to a learner's cognitive stage \citep{tao2025fivetiered}, and Load Reduction Instruction has been bridged explicitly to AI, tying cognitive-load theory to tool placement \citep{martin2025integrating}. Other work connects AI to tutoring, pedagogy, and expertise: a transformer-based Socratic tutoring system \citep{zhang2024spl} and a MOOC-to-AI unified pedagogy \citep{yuan2025bridging} each join cognition to AI, while Klein and Klein read expertise reversal together with cognitive atrophy to explain why the same tool can level a class or hollow it out \citep{klein2025extended}.

Two recent papers move especially close to the grain of learning activity itself. Vendrell and Johnston propose a design-oriented framework for scaffolding critical thinking with generative AI, built around cognitive friction, provisional AI partnership, embedded evaluation, metacognitive regulation, epistemic agency, and the sequencing of AI-mediated with AI-free phases \citep{vendrell2026scaffolding}. EFFORT-AI goes further toward phase-level instructional sequencing. It proposes a six-phase effort-preserving module (Elicit, Formulate, Feedback, Organize, Reflect, Transfer) with the rule that AI should preserve target cognition, require bounded productive effort before strong assistance, and adapt support to learner expertise, task complexity, and calibration \citep{liubertaite2026effortai}.

This convergence clarifies the contribution of the present frame. We offer a teacher-facing placement frame for the ordinary instructional problem of helping a student learn one idea. The frame separates the first hard attempt, the scaffolded middle, and the final unaided check: Prime, Probe, Point, Attach, Strengthen, and Test. It brings together cognitive-load placement, the social-motivational conditions that make productive struggle tolerable, and a practical rule teachers can carry into lesson design: protect Probe and Test, use guarded support in between, and treat effortlessness as the warning signal. If letting AI in makes the learning task feel effortless, it is probably replacing the work that should build the skill. The contribution is not to replace existing frameworks, but to translate their shared insight into a form teachers can use at the level of a single idea, task, or lesson. We make that translation concrete through paired classical and AI-supported menus, a secured AI-free check that keeps the proof the student's own, and a worked knowledge-map example of the learning process.

\section{The learning model: the six moves}
\label{sec:model}

\begin{figure}[t]
\centering
\includegraphics[width=\textwidth,height=0.85\textheight,keepaspectratio]{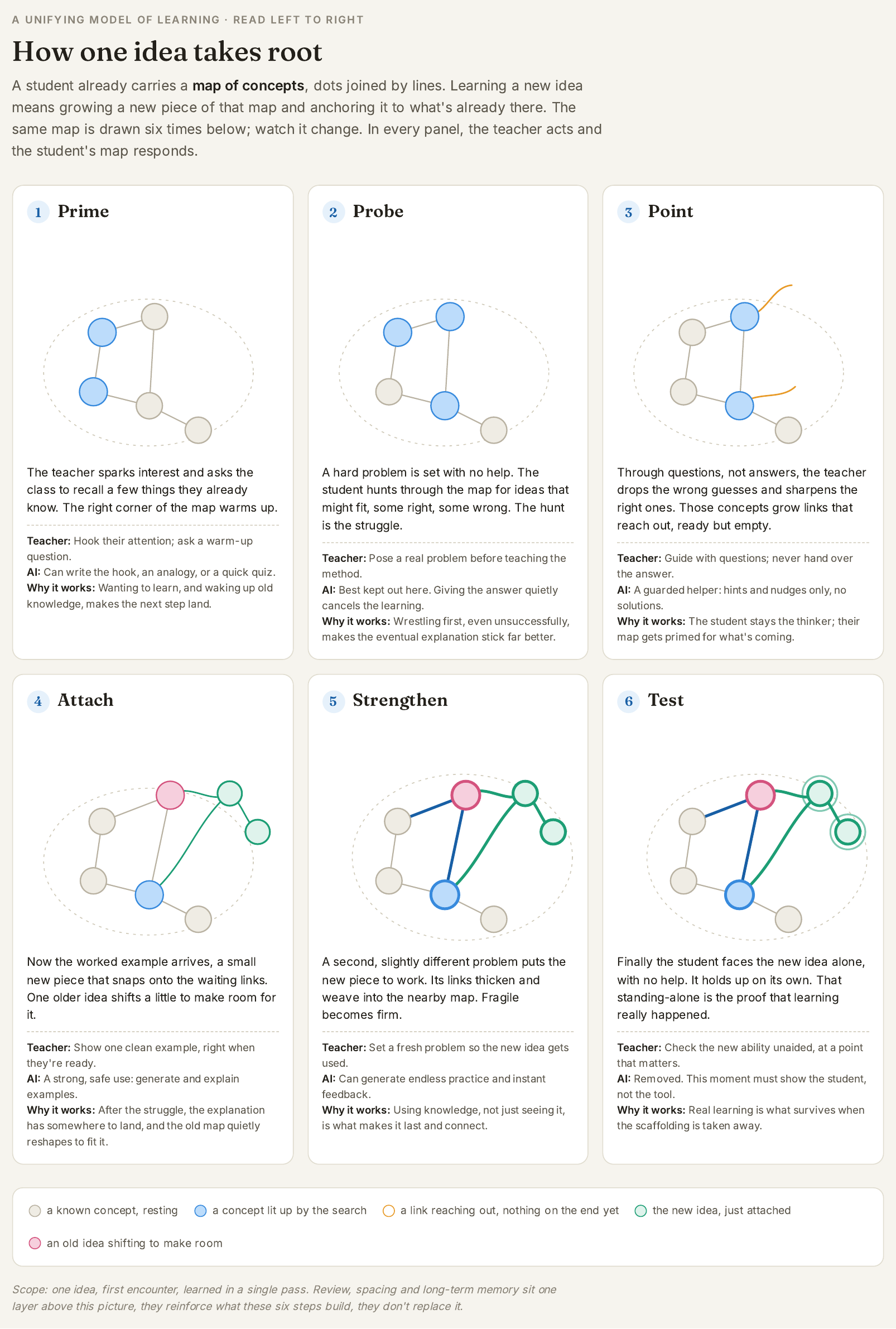}
\caption{The six moves, shown on the knowledge-graph metaphor. A new dot is primed, probed for, pointed toward, attached, strengthened, and finally tested unaided.}
\label{fig:phases}
\end{figure}

Picture what a student knows as a map: dots are ideas, and lines are the links between them. Learning one new idea means growing a new dot and tying it firmly into the map. That happens through six moves, in order. We call them \textbf{the six moves}: Prime, Probe, Point, Attach, Strengthen, Test, shown on Figure~\ref{fig:phases}. The figure marks, for each move, what the teacher does, what AI may or may not do, and how the student's knowledge map changes. Probe and Test are the two protected moments: the first hard attempt and the final unaided check, where the student's own work must remain visible.

\begin{enumerate}
  \item \textbf{Prime.} Before anything hard, the teacher sparks a reason to care and wakes up the prior knowledge the new idea will attach to. On the map, a few dots begin to glow. This is where belonging and high expectations do their work, the social layer that decides whether a student will risk the struggle that comes next \citep{walton2011,roorda2011}.
  \item \textbf{Probe.} The teacher sets a genuinely hard problem before teaching the method and holds back the answer. The student searches the map, trying ideas, many of them wrong. Being a bit stuck here is not wasted time; it is the learning beginning. This is productive failure, problem-solving before instruction, with measured benefit ($d \approx 0.36$ rising to about $0.58$ at high design fidelity) \citep{kapur2008,sinhakapur2021}. This is what R\'atz was doing when he set the problem before the rule.
  \item \textbf{Point.} The teacher now guides with questions, not answers. Wrong guesses fall away; promising ideas sharpen and reach outward, ready to grab the new idea the moment it arrives. This is Socratic guidance and cognitive apprenticeship, expert thinking made visible \citep{collins1989}.
  \item \textbf{Attach.} Only now, after the struggle has prepared the ground, does the teacher show one clean worked example. It snaps onto the waiting links, and one idea the student already held shifts a little to make room. The worked-example effect is strongest exactly here, for novices and after a real attempt \citep{swellercooper1985,sweller2011}.
  \item \textbf{Strengthen.} A new connection is fragile, so the student uses the idea on a second, varied problem. Each use thickens the links and weaves the new piece into the map. Deliberate practice, varied and interleaved, and consolidation by explaining it, including teaching it to a peer, live here \citep{dunlosky2013,chi2009,cohenkulik1982,roscoechi2007}. Teaching the idea to someone else is what Szubartowski's peer-tutoring leaned on.
  \item \textbf{Test.} Finally the student meets the idea with no help at all. If it holds on their own, learning has happened. That ability to stand without the scaffolding is the only proof that counts, and retrieval from memory is itself one of the strongest ways to cement it \citep{roediger2006,dunlosky2013}.
\end{enumerate}

The scope of this model is one idea, on first encounter, in a single pass. Long-term retention, which needs review and spacing, is a layer above it that reinforces, but does not replace, these six moves.

\section{The placement rule}
\label{sec:rule}

The rule is per skill, not per course. The same student, in the same hour, may need the tool kept out of one skill and welcomed into another. AI has two legitimate jobs. First, it cuts busywork that is not the point, so the student's effort goes to the real thinking. Second, it adds feedback and practice density that no single teacher could supply at scale. What it must not do is the one thing that is the point: the student's own struggle on the skill being learned, and the unaided performance that proves the skill is theirs.

That gives one rule. \textbf{AI belongs wherever it increases feedback, practice density, or professional realism without obscuring the evidence that the student can think and perform unaided.} Secure the first hard attempt and the final unaided check; allow guarded AI for feedback, examples, and drill in between; open AI for authentic work once unaided competence is shown. The diagnostic a teacher can carry into any task is short: \textbf{if letting AI in makes the task feel effortless, it is in the wrong place.} Effortlessness is the symptom of the tool doing the learning. Read it precisely, though: the point is not effort for its own sake. Busywork is effortful too, and it teaches nothing. What matters is effort on the skill being built, so AI belongs wherever it clears away the effort that is not the skill, the looking-up, the formatting, the dead ends, and must stay out wherever it would do the part that is. This is also the line the cognitive-load and expertise-reversal literature predicts: support that helps a novice becomes redundant, even harmful, for someone further along, so the same scaffold must fade \citep{kalyuga2007,kirschner2006}.

Mapped onto the six moves, the access pattern is: protected attempt, guarded support, protected proof. Probe and Test are AI-out moments: the first hard attempt and the final unaided check. Point, Attach, and Strengthen are guarded-support moments, where AI may add hints, examples, feedback, and practice without taking over the skill. Prime remains low risk because it comes before the struggle: AI may help generate hooks, analogies, or recall prompts, but it is not yet doing the skill-defining work.

Read at the level of a course or a programme, the same rule is a governance principle, not only a teaching one. An institution does not need a blanket position on AI, allow or ban; it needs a placement rule it can defend per skill. Stated this way, an AI-use policy stops being a prohibition list and becomes a design principle a department can adopt, a syllabus can declare, and a student can understand. The load-bearing point of such a policy is the secured check. When AI can complete most unsupervised work, the one tool-free point in the sequence, Test, is where the credibility of a grade, and of the credential behind it, is defended; governing that point well is as much a question of assessment integrity as of teaching \citep{dawson2021,bearman2024,teqsa2023}.

Table~\ref{tab:placement} states the rule move by move: the teacher's job, AI's job, where AI must step back, and the evidence behind each.

\begin{table}[ht]
\centering
\small
\caption{The placement map. For each move: the teacher's role, AI's legitimate role, and the supporting evidence. The two AI-out moves, Probe and Test, are the non-negotiable core of the rule.}
\label{tab:placement}
\begin{tabularx}{\textwidth}{L{0.10\textwidth} L{0.26\textwidth} L{0.40\textwidth} L{0.14\textwidth}}
\toprule
\textbf{Move} & \textbf{Teacher's role} & \textbf{AI's role (and the line)} & \textbf{Evidence} \\
\midrule[\heavyrulewidth]
Prime & Spark interest; wake prior knowledge; signal belonging. & Draft hooks; auto-generate a recall warm-up. Low risk; no struggle touched yet. & \citep{walton2011,dunlosky2013} \\
\rowsep
Probe & Set a hard problem; withhold the answer; protect the struggle. & \textbf{Stay out.} At most an access-timing gate that unlocks the tool only after an independent attempt. & \citep{kapur2008,sinhakapur2021,rotter2026} \\
\rowsep
Point & Guide with questions; give the smallest hint, then fade. & A guarded tutor: hints, error-spotting, or a \emph{related} solved problem to cue transfer, never the finished answer. The single most valuable placement. & \citep{bastani2025,collins1989,gickholyoak1983,wangcopilot2024} \\
\rowsep
Attach & Show one clean example, exactly when ready; name what shifts. & Generate tailored worked examples and re-framings on demand. A safe, strong use. & \citep{swellercooper1985,kestin2025} \\
\rowsep
Strengthen & Set a second, varied problem; have students explain and teach it. & Scale practice: calibrated drills, instant feedback, an AI tutee to teach. Keep the student doing the work. & \citep{dunlosky2013,chi2009,roscoechi2007} \\
\rowsep
Test & Step back; run a fair, secured, unaided check. & \textbf{Stay out of delivery.} May draft items for the teacher to vet; the check itself is AI-free. & \citep{roediger2006,dawson2021,bearman2024} \\
\bottomrule
\end{tabularx}
\end{table}

\section{Ideas: where AI could be put to work, move by move}
\label{sec:ideas}

This section sketches the option space before the full menus. The first menu is classical teaching, everything good teachers have always done with no technology required. The second is the AI menu, newer tools that can stretch or speed up the same moves. For most moves an AI tool does not replace a classical one; it scales it (more practice, faster feedback, for every student at once) or frees the teacher's scarce time for judgement, encouragement, and the protected struggle. The full menus are in Section~\ref{sec:menus}; this section gives the shape.

The clearest AI wins cluster where the move is already safe for scaffolding. In \textbf{Prime}, an AI-generated recall quiz scales the cheapest high-value move in the lesson, waking prior knowledge for every student at once. In \textbf{Point}, an answer-withholding tutor scales Socratic questioning, the placement the Bastani guardrail arm vindicates \citep{bastani2025}; it is the Socratic move a master like Aristotle once gave a single prince, now within reach of a whole room at once. AI can also find a related problem the student has already solved to cue analogical transfer \citep{gickholyoak1983}, or coach the human tutor rather than the student \citep{wangcopilot2024}. In \textbf{Attach}, on-demand worked examples are one of AI's strongest and lowest-risk uses \citep{swellercooper1985}. In \textbf{Strengthen}, infinite calibrated drills and instant formative feedback scale deliberate practice past what any teacher can hand-produce \citep{dunlosky2013}, and a student can even teach an AI playing a confused learner, a thin-evidence but promising stand-in for peer tutoring.

The master-teacher examples make the access argument concrete. The master teachers built problem-rich environments by hand, a steady bank of calibrated hard problems that needed a rare curator. AI's clearest, best-evidenced win is making that environment cheap and universal: the supply and methodology problem that kept master tutoring scarce is exactly the one AI is built to solve.

The two moves where AI must mostly step back are \textbf{Probe} and \textbf{Test}. In Probe, the only safe AI presence is an access-timing gate that unlocks the tool only after an independent attempt, which mirrors productive failure \citep{rotter2026}. In Test, AI may help draft items, but the delivery of the secured check stays AI-free, because that AI-free point is the whole evidence that the student can think without the tool \citep{dawson2021}. The pattern is consistent with the broader evidence: structured, answer-withholding use tends to help, while unstructured answer-giving can harm, and cognitive offloading, reducing internal effort by leaning on an external aid, improves immediate performance while breeding dependence \citep{risko2016}, and the AI-specific studies, though young and largely correlational, point the same way \citep{lee2025,fangasevic2025,gerlich2025}.

\section{The two menus}
\label{sec:menus}

These are the full menus behind Section~\ref{sec:ideas}. Effect strength is a rough guide from research: \estrong{} means repeatedly shown to work well; \estrongstar{} means the underlying teaching move is strong but the AI version of it still needs checking; \esolid{} means good support; \ectx{} means it depends heavily on how it is used; and \ereq{} marks a rule for a fair check, not an effect size. The research uses different kinds of evidence, so the rating is kept deliberately coarse.

\begin{longtable}{L{0.23\textwidth} c L{0.30\textwidth} L{0.27\textwidth}}
\caption{Menu A: what good teachers have always done (no AI needed).}\label{tab:menuA}\\
\toprule
\textbf{Technique} & \textbf{Effect} & \textbf{Why it works on the map} & \textbf{How it fits} \\
\midrule[\heavyrulewidth]
\endfirsthead
\toprule
\textbf{Technique} & \textbf{Effect} & \textbf{Why it works on the map} & \textbf{How it fits} \\
\midrule[\heavyrulewidth]
\endhead
\bottomrule
\endfoot
\phaserow{1}{Prime}{spark interest, wake prior knowledge}
Hook / curiosity gap & \ectx & Attention and motivation set how much effort is spent. & Sets up Probe; AI can draft hooks, teacher picks. \\
\rowsep
Activate prior knowledge & \estrong & Lights up the region the new idea must attach to. & Doubles as retrieval; cheapest high-value move. \\
\rowsep
Belonging \& high expectations & \esolid & Removes the fear that makes students disengage. & Underlies every step; protects the Probe. \\
\midrule
\phaserow{2}{Probe}{protected struggle before any telling}
Productive failure & \estrong & The search primes the map so the explanation lands deeply. & Core of the phase; keep answer-giving AI out. \\
\rowsep
Open-ended challenge & \esolid & Working at the edge of ability is where growth happens. & Calibrate difficulty; peer work shares the load. \\
\rowsep
Think-pair-share & \esolid & Every student generates an attempt before hearing others. & Makes struggle universal; leads into Point. \\
\midrule
\phaserow{3}{Point}{guide with questions, withhold the answer}
Questioning, not telling & \estrong & Keeps the student as the thinker. & Heart of the phase; the role to engineer AI into. \\
\rowsep
Hints \& scaffolding (then fade) & \estrong & Just enough support to keep struggle productive. & Fading is essential; AI can scale graduated hints. \\
\rowsep
Formative feedback & \estrong & Keeps the search from going blind; repairs the weak link. & Point, not solve; becomes the correction loop in Strengthen. \\
\rowsep
Peer instruction & \estrong & Explaining to a peer exposes and repairs gaps. & Mazur's method; works in large classes. \\
\midrule
\phaserow{4}{Attach}{one clean example, exactly when ready}
Worked example & \estrong & A clear example snaps in as the missing piece. & Best after a real attempt; AI generates these well. \\
\rowsep
Explicit instruction / consolidation & \estrong & Turns the messy attempts into the clean, usable rule. & The teaching half of productive failure. \\
\rowsep
Think aloud as an expert & \esolid & Shows the moves of thought, not just the answer. & Model, coach, fade; richer but slower. \\
\rowsep
Name what shifts & \ectx & Learning often re-shapes an old idea, not just adds one. & Prevents the new idea sitting inert. \\
\midrule
\phaserow{5}{Strengthen}{use it, vary it, weave it in}
Deliberate practice & \estrong & Repeated use thickens the new links to automaticity. & AI scales this; human diagnoses the weakness. \\
\rowsep
Mixed practice & \esolid & Forces the student to choose which idea applies. & Feels harder, teaches more; pairs with spacing. \\
\rowsep
Spaced / distributed practice & \estrong & Rebuilding the link after a gap makes it durable. & Lives above one lesson; Strengthen is its home. \\
\rowsep
Self-explanation & \esolid & Generating the explanation forces deeper integration. & Cheap and powerful; AI can prompt at scale. \\
\rowsep
Peer tutoring & \estrong & Teaching forces the tutor to organise and justify. & Szubartowski's lever; AI can stand in as the learner. \\
\midrule
\phaserow{6}{Test}{unaided, fair, real}
Recall / testing & \estrong & Pulling knowledge out cements it. & Doubles as learning and evidence; the test stays AI-free. \\
\rowsep
Oral / defence / explain-back & \esolid & Real-time questioning is hard to fake. & The most AI-proof check; use at meaningful points. \\
\rowsep
Process evidence & \ectx & Shows the thinking happened. & Complements, does not replace, the unaided check. \\
\end{longtable}

\begin{longtable}{L{0.23\textwidth} c L{0.30\textwidth} L{0.27\textwidth}}
\caption{Menu B: AI tools, mapped to the same six moves.}\label{tab:menuB}\\
\toprule
\textbf{AI technique} & \textbf{Effect} & \textbf{Why it works (and the catch)} & \textbf{Substitutes / complements} \\
\midrule[\heavyrulewidth]
\endfirsthead
\toprule
\textbf{AI technique} & \textbf{Effect} & \textbf{Why it works (and the catch)} & \textbf{Substitutes / complements} \\
\midrule[\heavyrulewidth]
\endhead
\bottomrule
\endfoot
\phaserow{1}{Prime}{generate the spark (low risk)}
AI-generated hook \& analogy & \ectx & Lifts motivation cheaply; pick what resonates. & Complements the human hook; no struggle touched. \\
\rowsep
AI-generated recall quiz & \estrongstar & Wakes prior knowledge for every student; items need checking. & Scales the cheapest high-value move; substitutes prep, not the recall. \\
\midrule
\phaserow{2}{Probe}{mostly keep AI out}
Access-timing gate & \esolid & Unlocks AI only after an independent attempt; mirrors productive failure. & The only safe AI presence here; turns a struggle-killer into a struggle-respecter. \\
\midrule
\phaserow{3}{Point}{the sweet spot, a guarded Socratic tutor}
Answer-withholding AI tutor & \estrong & Hints, not answers, removed the harm of answer-giving. & Scales Socratic questioning; the top placement. \\
\rowsep
Error-spotting \& misconception feedback & \esolid & Targets the individual's gap; must not solve it. & Covers students one teacher cannot reach in the moment. \\
\rowsep
Related prior problem & \ectx & Cueing a similar solved case can aid analogical transfer; must not show this solution. & A guarded hint; the target problem stays the student's own. \\
\rowsep
AI helping the human tutor & \esolid & Raises the floor of human feedback quality. & AI coaches the coach; cheap (\$20/tutor/year). \\
\midrule
\phaserow{4}{Attach}{a safe, strong generator}
On-demand worked examples & \estrongstar & Unlimited tailored examples; strong when correct and well-timed. & Scales the worked-example effect; teacher times it. \\
\rowsep
Alternative explanations / re-framings & \esolid & Finds the framing that clicks for a given student. & Complements the teacher's single explanation. \\
\midrule
\phaserow{5}{Strengthen}{scale practice and feedback}
Targeted practice sets & \estrongstar & Practice at the right level, density no teacher could hand-produce. & Scales deliberate practice; teacher diagnoses, student does the work. \\
\rowsep
Instant formative feedback & \esolid & Tightens the practice loop; faster correction. & Complements human feedback on volume and timing. \\
\rowsep
Spaced-review scheduling & \esolid & Automates high-value spacing; review must use recall. & Bridges the lesson to long-term retention. \\
\rowsep
Teach-the-AI & \ectx & Recreates the protege effect; student stays the teacher. & Scales peer tutoring; evidence still early. \\
\midrule
\phaserow{6}{Test}{AI steps back, secure the moment}
AI-generated assessment items & \ectx & Saves prep; delivery itself must be AI-free. & Complements design; the secured check stays human-run. \\
\rowsep
AI-assisted process review & \ectx & Makes process evidence practical at scale; a support only. & Complements oral/secured checks. \\
\rowsep
Removed at the unaided check & \ereq & One AI-free point shows the student can think alone. & Non-negotiable; everything else readies the student for it. \\
\end{longtable}

\section{A worked example: survivorship bias}
\label{sec:worked}

\begin{figure}[H]
\centering
\includegraphics[width=\textwidth,height=0.85\textheight,keepaspectratio]{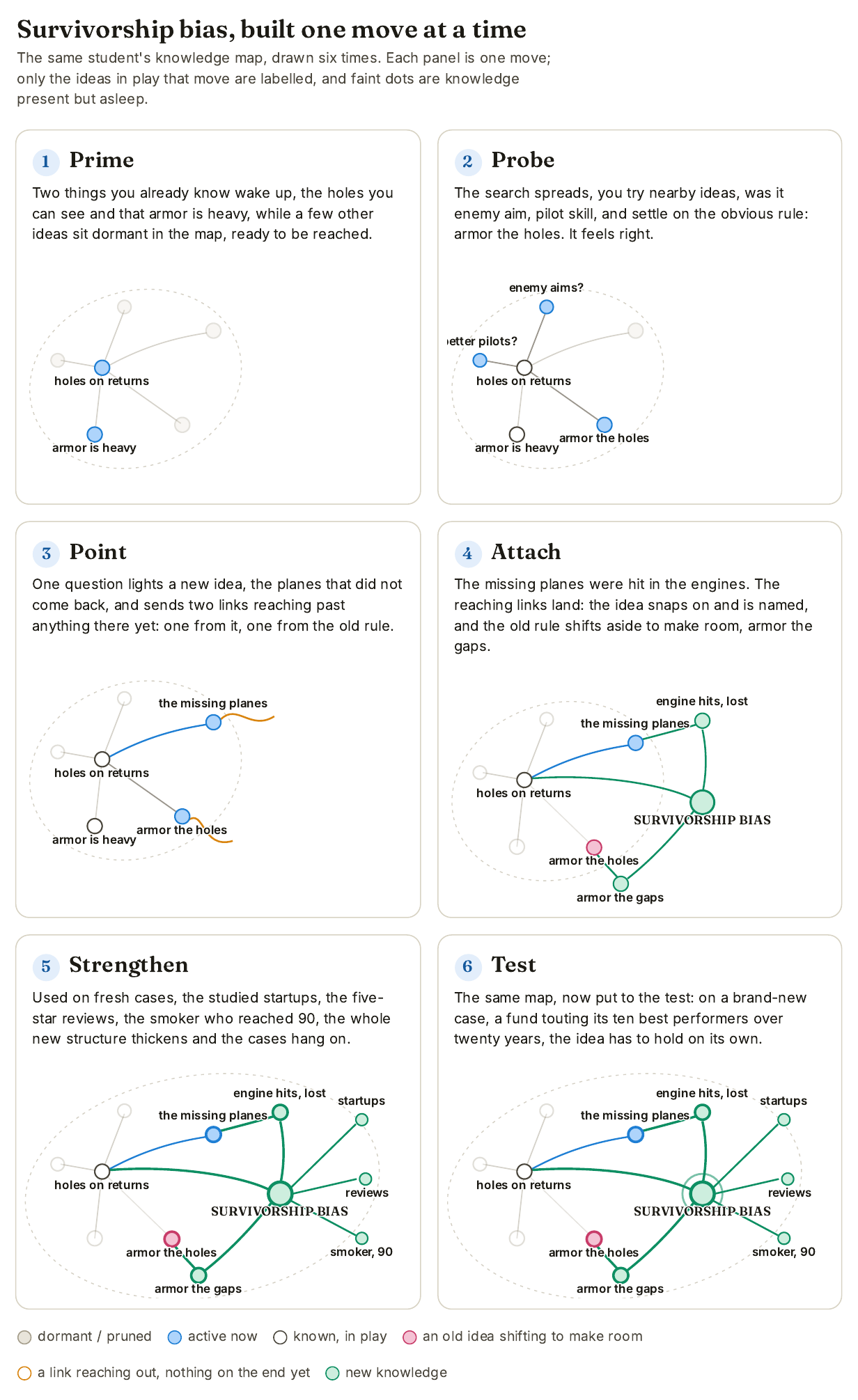}
\caption{Survivorship bias as a student's knowledge map across the six moves, growing inside a dashed field of knowledge.}
\label{fig:worked}
\end{figure}

To make the frame concrete, take one real idea and run it through the six moves. Survivorship bias is a good test case precisely because almost everyone forms the wrong rule first, which is the productive error the frame is built around. Table~\ref{tab:worked} gives the same lesson from the teacher's side, where the only column that carries the whole rule is the last: where AI is let in. Figure~\ref{fig:worked} shows the idea as a growing knowledge map, one panel per move.

\begin{table}[ht]
\centering
\small
\caption{The same worked example as a lesson plan. AI is out only at Probe and Test, the one struggle and the one proof that must be the student's own.}
\label{tab:worked}
\begin{tabularx}{\textwidth}{L{0.10\textwidth} L{0.50\textwidth} L{0.30\textwidth}}
\toprule
\textbf{Move} & \textbf{What the teacher does} & \textbf{Where AI goes} \\
\midrule[\heavyrulewidth]
Prime & Pose the 1943 puzzle: bombers return riddled with holes; where would you add armor? & \textbf{In.} No struggle yet; draft the hook. \\
\rowsep
Probe & Make the student commit first; almost everyone answers ``armor the holes.'' Withhold help. & \textbf{Out.} This search is what builds the skill. \\
\rowsep
Point & Ask, do not answer: what about the planes that did not return? & \textbf{Gated.} The same question or a related solved case, never the conclusion. \\
\rowsep
Attach & Reveal the planes hit in the engines; name survivorship bias; flip the rule to ``armor the gaps.'' & \textbf{In.} Generate the clean statement and a diagram. \\
\rowsep
Strengthen & Three fresh cases to classify (the studied startups, five-star reviews, the smoker who reached 90), each hiding the same missing data. & \textbf{In.} Scale varied practice and instant feedback. \\
\rowsep
Test & A fund advertises its ten best funds over twenty years; the losers were quietly closed and dropped. What is wrong with the claim? & \textbf{Out.} Standing alone is the proof. \\
\bottomrule
\end{tabularx}
\end{table}

\clearpage

In Figure \ref{fig:worked}, faint dots are ideas present but dormant. At Probe the search lights nearby ideas and settles on the natural wrong rule (armor the holes); Point lights a new idea (the missing planes) with links reaching out into empty space; Attach names the concept and shifts the old rule aside to ``armor the gaps''; Strengthen attaches fresh cases, thickens the links, and expands the field to hold them; Test re-examines the same map with the idea now standing on its own.

\section{Putting it to work}
\label{sec:adoption}

Adopting this does not mean redesigning a course. The smallest useful step is one lesson: take a single skill, hold the tool out of the first attempt and the final check, and let it scaffold the middle. The menus in Section~\ref{sec:menus} give a set of tools; a teacher picks the one or two moves where AI buys back the most time and leaves the rest. One lesson run this way teaches the frame fast and it is reversible, so the cost of trying is low.

At the level of a programme, the six moves give a faculty a shared vocabulary for redesign: a common map of where AI is licensed and where it is held back, consistent across the courses a student takes in sequence. This is the ground floor a redesign builds on. The department that agrees on the placement rule has already done the hardest part of the coordination that constructive alignment then formalizes \citep{biggs1996}. We offer the frame in that spirit, as a usable basis for whoever finds it helpful.

\section{Limitations, scope, and honest caveats}
\label{sec:limits}

\paragraph{Frame.} This is the frame that covers only how a student learns one idea and where AI fits that arc. It deliberately stops short of the operational layer: the assessment architecture, the AI-use contracts, the syllabus templates, and the per-course blueprints. These are out of scope here.

\paragraph{Scope of the model.} The six moves describe one idea, on first encounter, in a single pass. Durable retention needs review and spacing on top of this, a second layer that reinforces but does not replace the moves here \citep{dunlosky2013,roediger2006}. The sequence also assumes what it cannot itself supply: a motivated, engaged student. Securing that, and designing the experience around it, is the educator's craft, the part AI amplifies but does not replace, the teacher pacing a lesson the way a game designer paces a player, a chef sequences a tasting menu, or a guide leads a traveller through a country worth seeing.

\paragraph{The AI evidence is young.} The AI-specific layer (2023--2026) is fast-moving, heterogeneous, and partly preprint. The settled cognitive and social science carries the argumentative weight; the AI studies show that design decides the sign of the effect, not a fixed magnitude. We flag preprints in text: the human-AI tutoring result \citep{wangcopilot2024}, the access-timing study (a single lab, $N=105$) \citep{rotter2026}, and the cognitive-offloading EEG work \citep{kosmyna2025}, which is small, contested, and must be read with its published critique \citep{stankovic2025}. The Kestin doubling comes from a single elite crossover course and should not be read as a universal magnitude \citep{kestin2025}.

\paragraph{Master-teacher cases are motivation, not proof.} The Aristotle, Keller, R\'atz, and Szubartowski stories illustrate what scarce one-to-one mentoring could achieve; they cannot establish causation. Selection and survivorship effects dominate, the figures come from non-peer-reviewed sources, and there is a documented dissenting view that such intensive programs can demotivate non-prodigies \citep{mistrz2025}. The causal bridge to "scale the tutor" is carried by the tutoring-effect literature, not the anecdotes \citep{bloom1984,vanlehn2011}.

\paragraph{What this is, and is not.} This paper is a practical synthesis, without new empirical results. The pieces come from established and emerging work: cognitive-load placement, productive struggle, social-motivational support, and guarded AI use. The aim is to organize them into a compact placement frame a teacher can apply to one idea, task, or lesson. Its warrant is therefore partly theoretical and partly practical: whether it helps educators decide where AI should support learning, where it should be constrained, and where it should stay out. Future empirical work should test the frame directly, including whether protecting Probe and Test improves delayed unaided performance compared with answer-first or fully open AI use.

\section{Conclusion}

The same AI tool can help a student build a skill or quietly do the building for them, and the difference is where in the learning it sits. A new idea is learned through six moves, Prime, Probe, Point, Attach, Strengthen, Test, and AI earns its place in some and endangers others. Secure the first hard attempt and the final unaided test, scaffold with guarded AI in between, and open the tool for authentic work once competence is shown. One diagnostic carries the whole frame: if letting AI in makes the task feel effortless, it is in the wrong place.

Placed this way, AI has a genuine chance to scale the patient one-to-one attention that was, for most of history, the privilege of a very few. Placed wrong, it removes the struggle that learning needs. The stakes run past any single grade: the durable, transferable competence a student earns through that struggle is also the individual ground of what one of us has called \emph{cognitive sovereignty}, the capacity for autonomous thought in an age of AI systems built to supply it on demand \citep{clark1998,brcic2025memorywars}. Outsourced at the moment of learning, that ground erodes one effortless task at a time. With the right pacing, a lesson can keep both the struggle that builds the skill and the boost that scales it. The choice is resolved in the design of the lesson, not in a usage policy.

\bibliographystyle{apalike}
\bibliography{refs}

\end{document}